# EPiT : A Software Testing Tool for Generation of Test Cases Automatically

Rosziati Ibrahim[#1], Ammar Aminuddin Bani Amin[#2], Sapiee Jamel[##3], Jahari Abdul Wahab[*4]

[#]*Department of Software Engineering, Universiti Tun Hussein Onn Malaysia, 86400, Malaysia*

[##]*Department of Information Security,, Universiti Tun Hussein Onn Malaysia, 86400, Malaysia*

[*]*Engineering R&D Department, Sena Traffic Systems Sdn. Bhd., Kuala Lumpur, 57000, Malaysia*

***Abstract*** — *Software test cases can be defined as a set of condition where a tester needs to test and determine that the System Under Test (SUT) satisfied with the expected result correctly. This paper discusses the optimization technique in generating cases automatically by using EpiT (Eclipse Plug-in Tool). EpiT is developed to optimize the generation of test cases from source code in order to reduce time used for conventional manually creating test cases. By using code smell functionality, EpiT helps to generate test cases automatically from Java programs by checking its line of code (LOC). The implementation of EpiT will also be presented based on several case studies conducted to show the optimization of the test cases generated. Based on the results presented, EpiT is proven to solve the problem for software tester to generate test case manually and check the optimization from the source code using code smell technique.*

## I. INTRODUCTION

There are many techniques for a software tester to conduct a SUT within testing phase. Typically, a software tester manually generates the test cases and testing each of the module. This technique however may take longer time where most of the activity of generating test cases are prone to redundancy [1]. Moreover, a software tester needs to be precisely validate the test cases well in order to avoid ambiguity presented in Software Test Plan documentation. In a nut shell, automation tool helps to increase the reliability of system while reducing the cost of manual software testing [2 - 4].

To solve the problem of manually generating test cases, EpiT can be used as a software testing tool where test cases can be automatically generated from the source code. By using the code smell functionality, the tool can generate test cases by examining line by line of the source codes. Thus, EpiT optimization will help to find and provide high quality solutions.

This paper discusses the optimization technique using code smell algorithm in order to reduce the generation of test cases due to redundancy of test cases as well as an approach to automate the process of generating the test cases. The rest of the paper is organized as follows. It consists of 7 Sections. Related works are discussed in Section 2 and Section 3 discuss the application involved. Meanwhile, Section 4 demonstrates the framework of EpiT. Section 5 provides the implementation of EpiT tool using code smell to generate test cases from source code. Section 6 shows the results after the implementation phase and conclusion is in Section 7.

## II. RELATED WORK

Research paper of Albert [5] presents jPET which is a white-box test-case generator (TCG). jPET is a software testing tool build on Eclipse environment which automatically generate test-cases from the bytecode compiled from the java class. In order to yield this information to the user at the source code level, jPET performs reverse engineering of the test-cases obtained at the bytecode level by PET. By using jPET as integrated Eclipse programming, software developers can directly conduct testing activities in development phase.

In the study by Zhenzhen Wang, and Qiaolian Liu [6] discuss an improved Particle Swarm Optimization (PSO) algorithm to generate test case automatically. This study evaluated several techniques such as PSO and Genetic algorithm to determine perfomance in term of generating test cases. Based on the result, shows that the improved PSO is effectively reduce run time in generating software test cases automatically compare to conventional PSO.

The study by Hanyu Pei et al. [7] proposed a cloud-based Dynamic Random Test (DRT) technique for handling the evaluation of test case prioritization and resource allocation. This study also evaluated several techniques between the cloud-DRT and other techniques such as Round Robin Schedule (RRS) and cloud-based Random Partition Testing (RPT). All these experiments of evaluation were conducted to find fault detection effectiveness. Based on the result, it was concluded that the proposed technique uses less test cases to detect all seeded faults compared to RRS and cloud-based RPT. AMOGA [8] is another study for software testing. AMOGA is used as a framework in generating test cases from GUI for mobile applications.

Yunqi Du et al. [9] proposed a hybrid techniques which merging a mutation testing technique with genetic algorithm. These combination technique will be able to design test case optimization methods. The method helps in producing the test cases generation. Based on the results, they are able to show their combination technique have better score in term of an average variation score of more than 95% on the variants of the complete set. Thus, this study proved that using mutation testing combine with genetic algorithm leads to higher coverage and mutation score within test cases.

Based on study by Annibale Panichella et al. [10] conducted on 346 Java classes to assess the performance of DynaMOSA compared with whole-suite approach (WS), its archive-based variant (WSA) and Many-Objective Sorting Algorithm (MOSA). By Using DynaMOSA technique, test case can be generated and maximize test coverage in SUT. Based on the experimental result, it shows that DynaMOSA improved 8 percent in average coverage compared to MOSA technique. Table 1 shows the summary of the related works discussed.

**TABLE I: SUMMARY OF RELATED WORK**

| No | Authors & Year | Techniques | Outcome |
|---|---|---|---|
| 1 | Elvira Albert,<br><br>(2011) [5] | jPET: an Automatic Test-Case Generator Plugin | Fully integrated test case generation within the software development process. |
| 2 | Zhenzhen Wang, Qiaolian Liu<br><br>(2018) [6] | Particle Swarm Optimization (PSO) algorithm | The technique proved that is better compared to conventional PSO. |
| 3 | Hanyu Pei et al.<br><br>(2018) [7] | Dynamic Random Testing Strategy (DRT) | The technique proposed more effective in generating test cases compared to RPT and RRS. |
| 4 | Salihu et al.<br><br>(2018) [8] | AMOGA | AMOGA is used as a framework for testing mobile applications by using GUI |
| 5 | Yunqi Du et al.<br><br>(2019) [9] | Hyrbid Technique merging Mutation Testing Techniques with Genetic Algorithm | The hybrid techniques leads to better coverage and better mutation score in generating test cases. |
| 6 | Annibale Panichella et al.<br><br>(2018) [10] | Dynamic Many-Objective Sorting Algorithm<br><br>(DynaMOSA), | The technique proved to more high coverage in test case generation compared with MOSA. |

## III. EPIT FRAMEWORK

In this section, the framework of EpiT is clearly defined to demonstrate how the generation of test cases are derived. Fig. 1 shows the framework for EPiT. Source code is the input for EPiT. Then, the parser will be used to read line by line of the source codes. EPiT will detect classes with code smell and then generate test cases automatically based on detected methods.

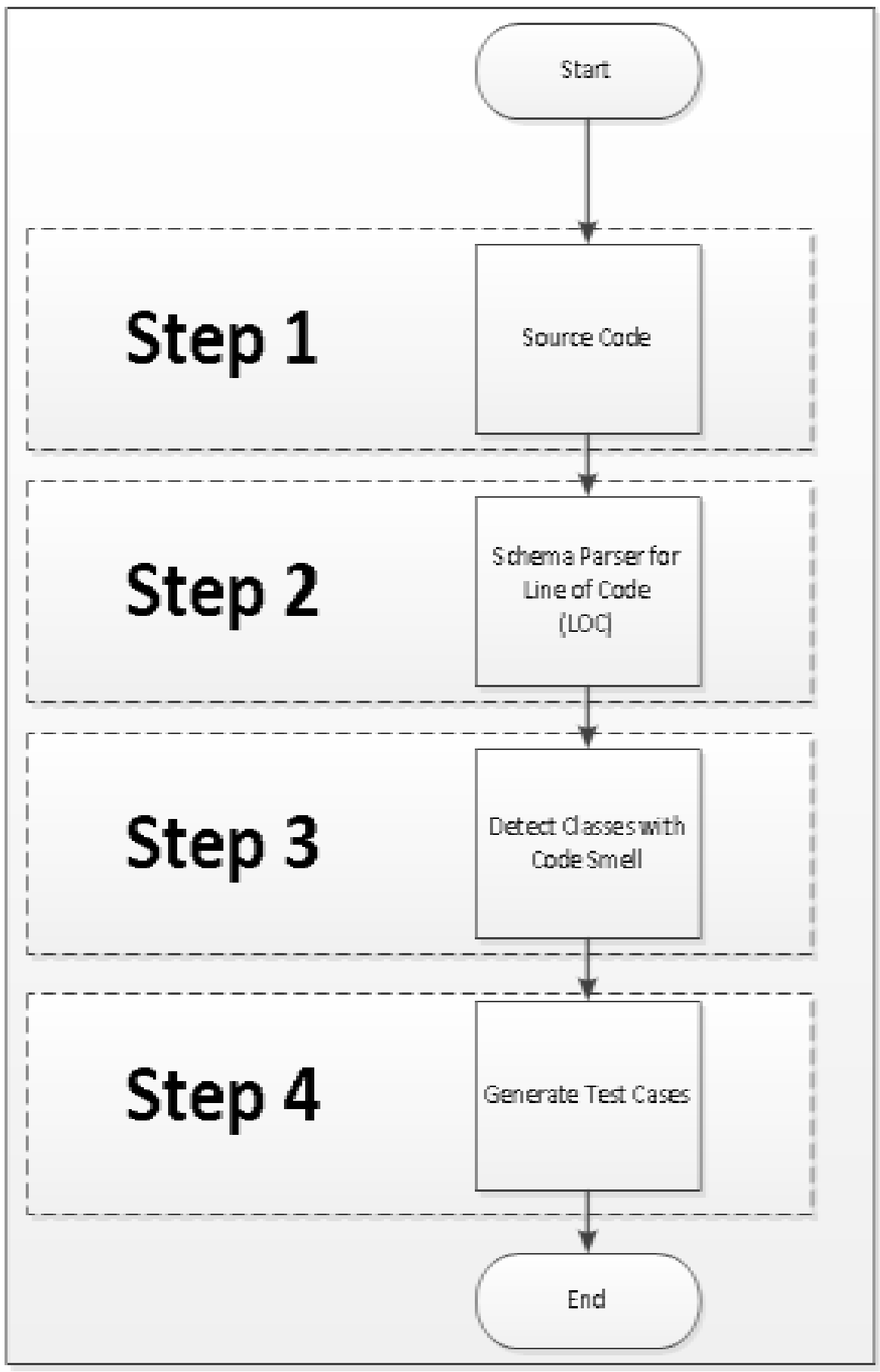

**Fig. 1: The Framework of EPiT**

Based from Fig. 1, a software tester can generate the test cases using EpiT in Eclipse Integrated Development Environment (IDE). Eclipse IDE is always been used to develop wide scope of application because of open source license. However, the development of EpiT for source code is currently limited to Java programming only. The details discussion for the framework has been discussed in [11].

## IV. THE CASE STUDIES

In this paper, several applications were implemented in order to test the Epit framework. All four of the applications were selected and downloaded from open source community Github.

The first application is the calendar application [12] which are used to view yearly and monthly calendar given input from user. Additionally, this application also calculates age where the application count the duration from birth of date to current year.

Secondly, the BlackJack application [13] is a card game where each player will dealt a card until a certain point. Player who get the highest hand win the round. The application consists of seven classes in total.

Thirdly, the CoffeeMaker application [14] is an inventory system for making a coffee. A user can add, edit or delete each coffee recipe to the inventory. Moreover, user can view all the listed recipes in the inventory. This application consists of four classes in total.

Lastly, the Elevator application [15] is an elevator system where allocate elevator cart to assign floor. Many functions were embed in the application such as the elevator cart will repositioned itself to the lowest floor when in idle state. This application also check sequence queue and sorted which floor should be stop first.

## V. IMPLEMENTATION OF EPIT TOOL

Base from the framework of EPiT in Fig. 1, the tool is implemented according to four steps. These steps are discussed in this section.

### A. Step 1: Source Code

Example of calendar case study using source code is shown in Fig. 2. The application consists of two methods in Java programming. Firstly, the project will be imported in Eclipse environment. Then the codes will be analyzed by EpiT by using the menu as shown in Fig. 3.

### B. Step 2: Schema Parser

By default, EpiT will read line by line according to the algorithm. Epit uses Java parser to get an Abstract Syntax Tree (AST). Based on Fig. 4, example of the AST code where is a structure representing Java programming in a way that it is easy to check and validate.

### C. Detect Classes with Code Smell

Epit will start to analyze and detect all the classes in the source using code smell functionality [16]. The parser will detect the class by node. Then within each of the node, EpiT identifies as method which has several other attributes such as method name, return type and input parameter. All the method classes detected will be stored in variable in EpiT. Fig. 5 shows example of code smell functionality in EpiT.

### D. Generate Test Cases

Then, test cases are generated based on the attributes that have been identify within the code smell functionality. Each of the test cases will have at least three possible scenario such as valid input, invalid input and null input. Example is shown in Fig. 6 where each of the test cases will be generated with multiple scenario.

```
/** Get the start day of the first day in a month */

static int getStartDay(int year, int month) {

  //Get total number of days since 1/1/1800 on tuesday
  int startDay1800 = 3;
  int totalNumberOfDays = getTotalNumberOfDays(year, month);

  //Return the start day
  return (totalNumberOfDays + startDay1800) % 7;
}

/** Get the total number of days since January 1, 1800 */

static int getTotalNumberOfDays(int year, int month) {
 int total = 0;

 //Get the total days from 1800 to year - 1
 for (int i = 1800; i < year; i++)
 if (isLeapYear(i))
   total = total + 366;
  else
   total = total + 365;

  //Add days from January to the month prior to the calendar month
  for (int i = 1; i < month; i++)
   total = total + getNumberOfDaysInMonth(year, i);

  return total;
}
```

**Fig. 2: Source Codes for Calender**

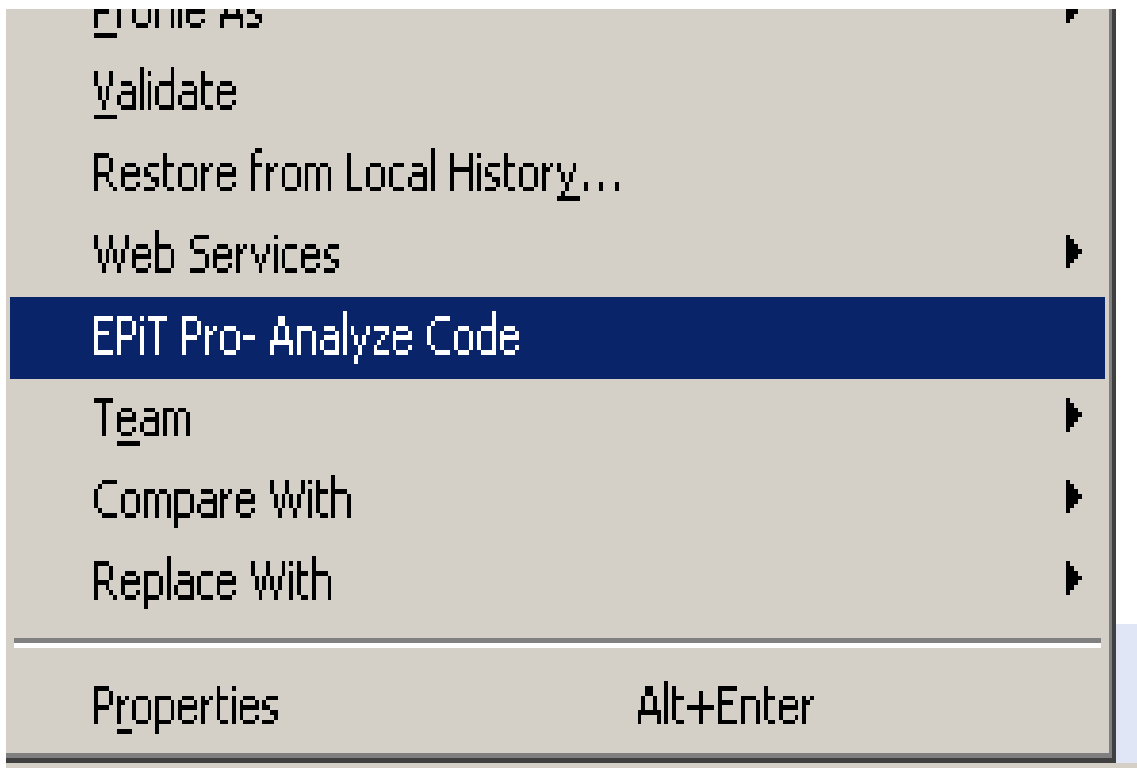

**Fig. 3: EPiT Analyzer**

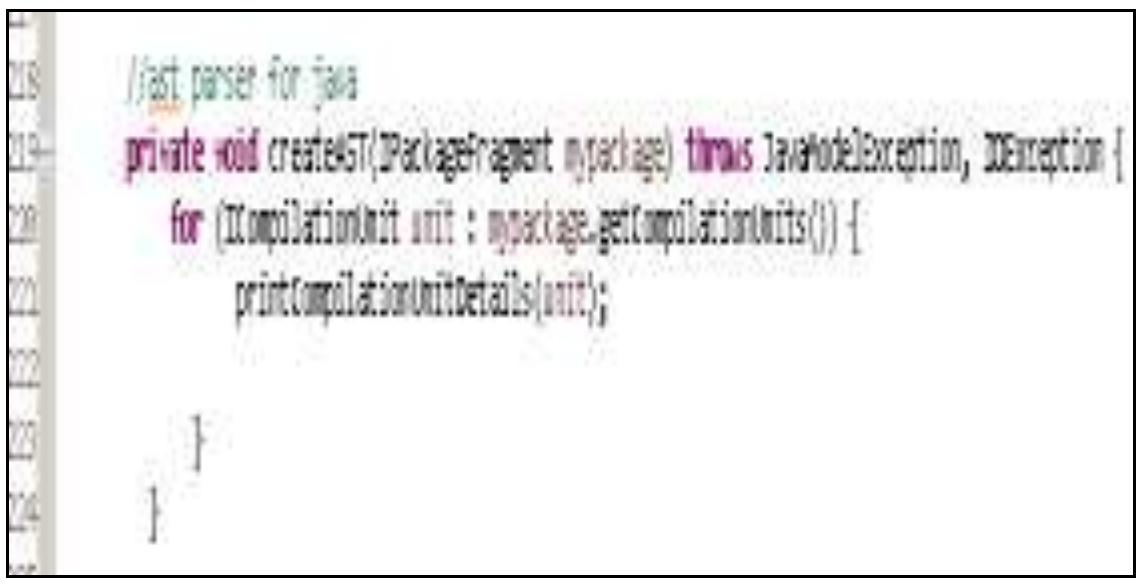

**Fig. 4: AST Code**

After EPiT has finished analyzed all the source code, it will prompt the success popup. Fig. 7 shows example of the popup with time duration taken on each process.

```java
private void printIMethodDetails(IType type) throws JavaModelException, IOException {
    IMethod[] methods = type.getMethods();
    for (IMethod method : methods) {

        String methodNameString="\nMethod name :" + method.getElementName()+"\n";
        Factory.getConsole().newOutputStream().write(methodNameString.getBytes());
        testCaseList.add(method.getElementName().toString());
        String signatureString="Signature :" + method.getSignature()+"\n";
        Factory.getConsole().newOutputStream().write(signatureString.getBytes());
        String returnTypeString="Return Type :" + method.getReturnType()+"\n";
        Factory.getConsole().newOutputStream().write(returnTypeString.getBytes());

        //get variable name in method
        String[] paramvalue=method.getParameterNames();
        String inputParamString="Input variable :"+"\n";
        Factory.getConsole().newOutputStream().write(inputParamString.getBytes());
        for(int p=0;p<paramvalue.length;p++){
            String resultParam=paramvalue[p].toString()+"\t";
            Factory.getConsole().newOutputStream().write(resultParam.getBytes());
        }
```

**Fig. 5: Code Smell in EPiT**

```java
//test scenario possible inputs
private String generateTestCase(String type,Integer boundary,String testMethodNameOrig,String[] paramvalue) {
    String result=null;
    if (type.equals("valid")){
        result="Test Case 1 : valid "+Arrays.deepToString(paramvalue)+"are input with :"+boundary;
        //System.out.print("//valid value\n    assertTrue(t."+testMethodNameOrig+"("+boundary+"));\n");
    }
    else if (type.equals("invalid")){
        result="\nTest Case 2 : invalid "+Arrays.deepToString(paramvalue)+"are input with :"+boundary;
        //System.out.println("//invalid value\n    assertFalse(t."+testMethodNameOrig+"("+boundary+"));\n");
    }
    else{
        result="\nTest Case 3 : null "+Arrays.deepToString(paramvalue)+"are input with :"+boundary;
        //System.out.println("//null value\n    assertFalse(t."+testMethodNameOrig+"("+boundary+"));\n");;
    }
    return result;
}
```

**Fig. 6: Generation of Test Cases**

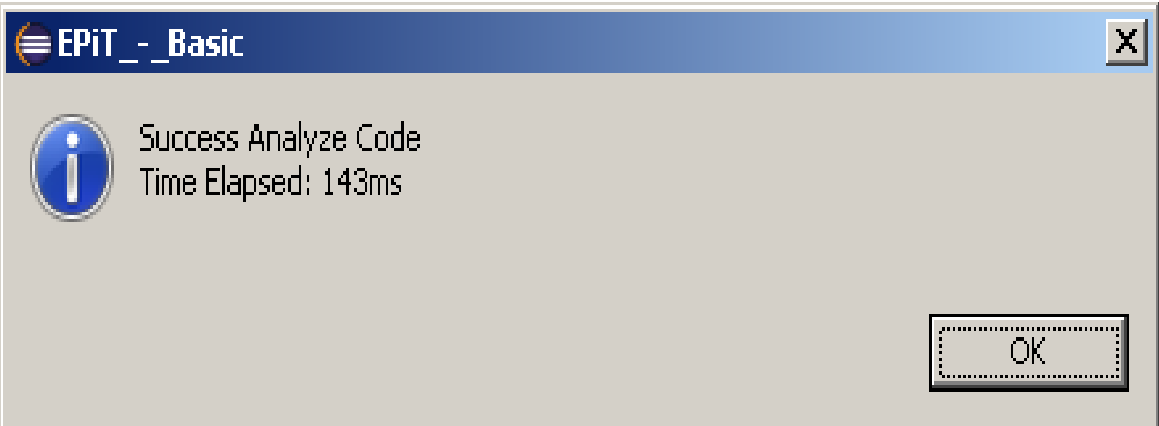

**Fig. 7: Success Dialog Box in EPiT**

Next, during analyzing the source codes, EPiT will output all details in the EpiT console. The console will show many attributes such as time and date of the conducted process, name of the project, and details each of test cases generated. Fig. 8 shows that the EpiT console of each of analyzed process.

Finally, EPiT will summarize all the analyzed source code and generated test cases in the EpiT console. The console will show project name, total generated test cases, duration of each process conducted. Fig. 9 shows that the EpiT summarization of each of successful process.

```
Console EPiT

Start Time: 2019/12/11 11:04:02
### Analyzing project :Algo Calendar Project ###

Details <Packages> :mypackage
-------------------------------------------------
Source file Menu.java

FullPath /Algo Calendar Project/src/mypackage/Menu.java
Has number of lines: 144
Details are :-

Method name :main
Signature :([QString;)V
Return Type :V
Input variable :
args
Generate Possible Test Scenario:
Test Case 1 : valid [args]are input with :1
Test Case 2 : invalid [args]are input with :-1
Test Case 3 : null [args]are input with :null

Method name :calculateAge
Signature :(QDate;)QCCalendar;
Return Type :QCCalendar;
Input variable :
mydob
Generate Possible Test Scenario:
Test Case 1 : valid [mydob]are input with :1
Test Case 2 : invalid [mydob]are input with :-1
Test Case 3 : null [mydob]are input with :null

-File analyzed End-
```

**Fig. 8: The Console for EPiT**

```
################ Summary ###################

Project Name :Algo Calendar Project
Total Package :2
Package Name :
        <mypackage>
        <mypackage>
Total Files :3
File Name :
        [Menu.java]
        [CCalendar.java]
        [CCalendarTest.java]
Total LOC :374.0
Total Test Cases Without Refactoring:15
Total Test Cases With Refactoring: 0
Optimization After Refactoring: 0%
JUnit BUilder: No

##############################################

Start Time: 2019/12/11 11:08:07
End Time: 2019/12/11 11:08:07
Time Elapsed: 82ms
```

**Fig. 9: Summary of EPiT Analyzer**

## VI. RESULTS AND DISCUSSIONS

Based on Table II, comparison result between each of case study is shown. The result shows that the calendar application is the fastest with 82ms and generated fifteen test cases. Meanwhile, the slowest is the coffee maker application which is 206ms with fifty seven test cases were generated.

The comparison had been led by considering one perspective. The differences of each application by duration time to generate test cases. From fastest time which is 81ms and the slowest time which is 206ms.

There are many factors which lead to inconsistent result which are project code complexity, CPU usage and memory usage. Although the result is not consistent, this project had proven that EpiT is faster than conventional manually generated test cases. Thus, those studies support the result of this project.

**TABLE II: RESULTS OF CASE STUDIES**

| *Application* | *Total LOC* | *Test Cases Generated* | *Duration Time (ms)* |
|---|---|---|---|
| Calendar | 374 | 15 | 82 |
| BlackJack | 1017 | 41 | 163 |
| CoffeeMaker | 982 | 57 | 206 |
| Elevator | 1183 | 66 | 149 |

## VII. CONCLUSIONS

This paper has discussed the works of the generating test cases using EpiT in Eclipse environment. Through the experiment, this research verified that EpiT is more reliable and faster response time compared to conventional manual testing technique. By analyzing the result, test cases can be optimized and help software tester to remove any redundancy in the source code. Based on the results, it is proven that using code smell functionality is a very efficient way to automatically generate test cases from source code. Other techniques for software testing can also be investigated and implemented such as mutation software testing technique, regression software testing technique and fuzzy software testing technique.

## ACKNOWLEDGMENT

The authors would like to thanks Ministry of Education (MOE) for supporting this study under Prototype Research Grant (PRGS) Vote No K037.